\documentclass[12pt]{iopart}
\usepackage{graphicx}

\usepackage{iopams}

\begin{document}

\title[]{A detailed description of the experimental realisation of quantum illumination protocol}

\author{ E.D.~Lopaeva$^{1,2}$, I.~Ruo Berchera $^1$, S.~Olivares $^{3,4}$, G.~Brida $^1$, I.P.~Degiovanni $^1$, M.~Genovese $^1$}

\address{$^1$ INRIM, Strada delle Cacce 91, I-10135 Torino, Italy}
 \address{$^2$ DISAT, Politecnico di Torino, I-10129 Torino, Italy}
\address{$^3$ Dipartimento di Fisica, Universit\`a degli Studi di Milano, I-20133  Milano,
Italy}
 \address{$^4$ CNISM UdR Milano
Statale, I-20133 Milano, Italy}

 \ead{m.genovese@inrim.it}

\begin{abstract}
In the last years the exploitation of specific properties of quantum states has disclosed the possibility
  of realising tasks beyond classical limits, creating the new field of quantum technologies
  \cite{4,5,6,7,8,9,10,11,prl2}. Among them, quantum metrology and imaging aim to improve the sensitivity and/or resolution of measurements exploiting non-classical features such as squeezing and quantum correlations (entanglement and discordant states) \cite{Kolobov,Treps:03,boy:08,bri:10, gio:11}. Nevertheless, in most of the realistic scenarios losses and noise are known to nullify the advantage of adopting quantum strategies \cite{Walmsley-PRL-2011}.
In this paper we describe in detail the first experimental realization of quantum illumination protocol aimed to target detection in a noisy environment, that preserves a strong advantage over the classical counterparts even in presence of large amount of noise and losses. The experiment, inspired by the theoretical ideas elaborated in \cite{llo:08,tan:08,sha:09,gua:09} (see also \cite{ms,msplus}), has been performed exploiting only photon number correlations in twin beams. Thus,  for its simplicity it can find widespread use. Even more important by challenging the common believe  that real application of quantum technologies is limited by their fragility to noise and losses, it paves the way to their real application.
\end{abstract}

\maketitle

\section{Introduction}
In our scheme \cite{prl} for target detection a probe beam of a bipartite correlated state may be partially reflected by an object
towards a camera, which also receives a thermal field acting
as noisy unknown background (thermal bath). \textbf{Our goal
is to investigate the performances of the quantum protocol, in a detection framework in which only photon numbers (i.e. intensities) are measured, with respect to the best classical
counterpart, namely a classically-correlated-light based protocol. We show as the use of
simple second order correlation measurements already suffices in
guaranteeing strong advantages to the quantum protocol.} This is a fundamental progress toward a practical realisation respect to some previous similar theoretical proposals
\cite{tan:08,sha:09,sha:PRA:09}, stemming from the ``quantum illumination'' scheme of \cite{llo:08}, where the discrimination strategy, based on quantum Chernoff bound
\cite{aud:07,cal:08}, was very challenging from an experimental point of view.
We
realise quantum target detection both by using quantum illumination (QI), specifically twin beams (TWB), and by using classical illumination (CI),
e.g. correlated thermal beams (THB)\textbf{, representing the best classical state in the specific detection framework,} pointing out unequivocally the experimental
advantage of the quantum protocol in mesoscopic regime, independently on the noise level.
The realisation of QI protocol, beyond paving the way to future practical application, also
provides a significant example of ancilla assisted quantum protocol
besides the few previous ones, e.g.\cite{bri:10,bri:ea,tak,al}. As a first application of quantum illumination to QKD, with a different detection scheme,  refer to the recent paper \cite{shap}.

\section{An experimental setup}

\begin{figure}[tbp]
\begin{center}
 \includegraphics[width=0.5\textwidth]{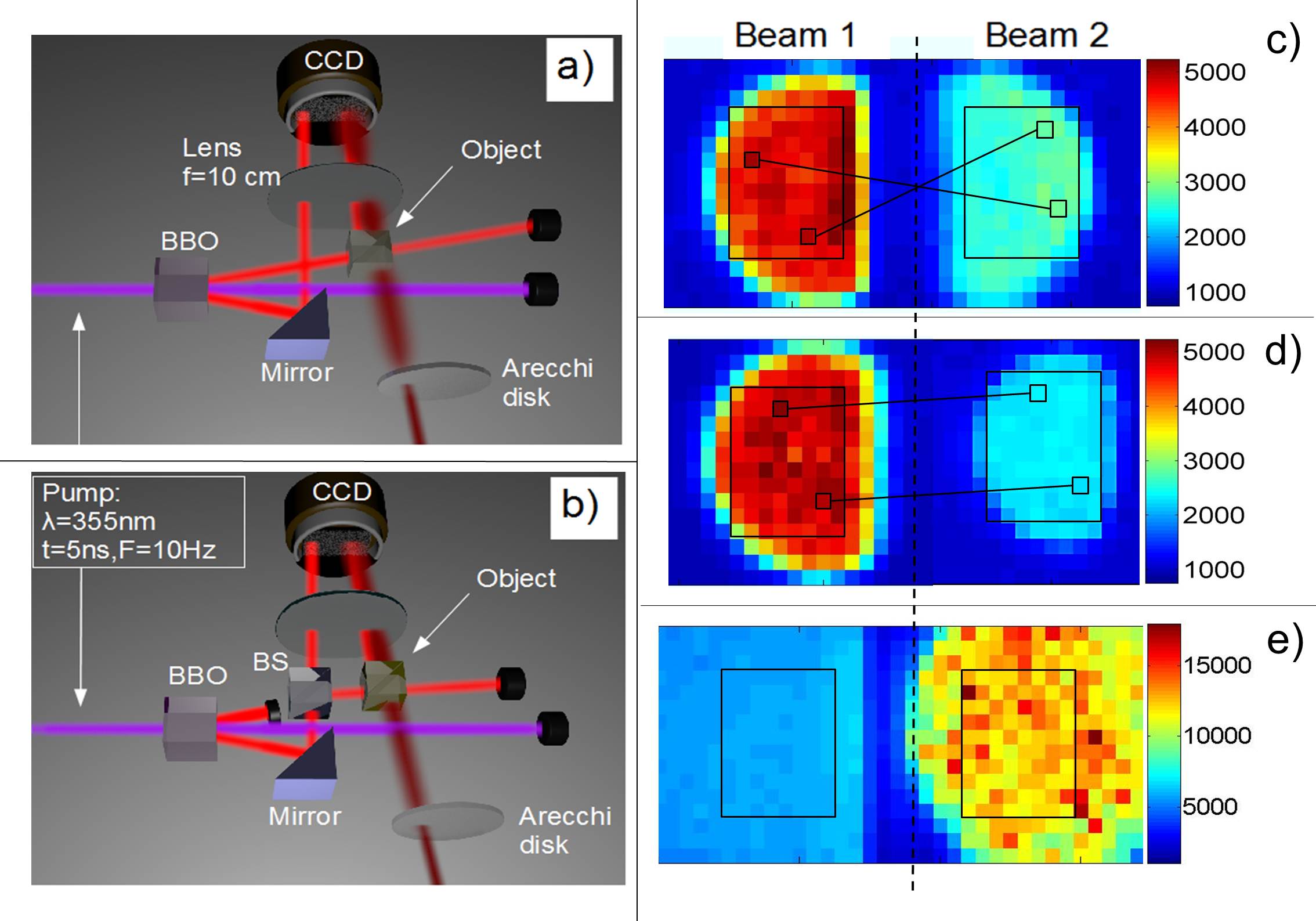}
\includegraphics[width=0.23\textwidth]{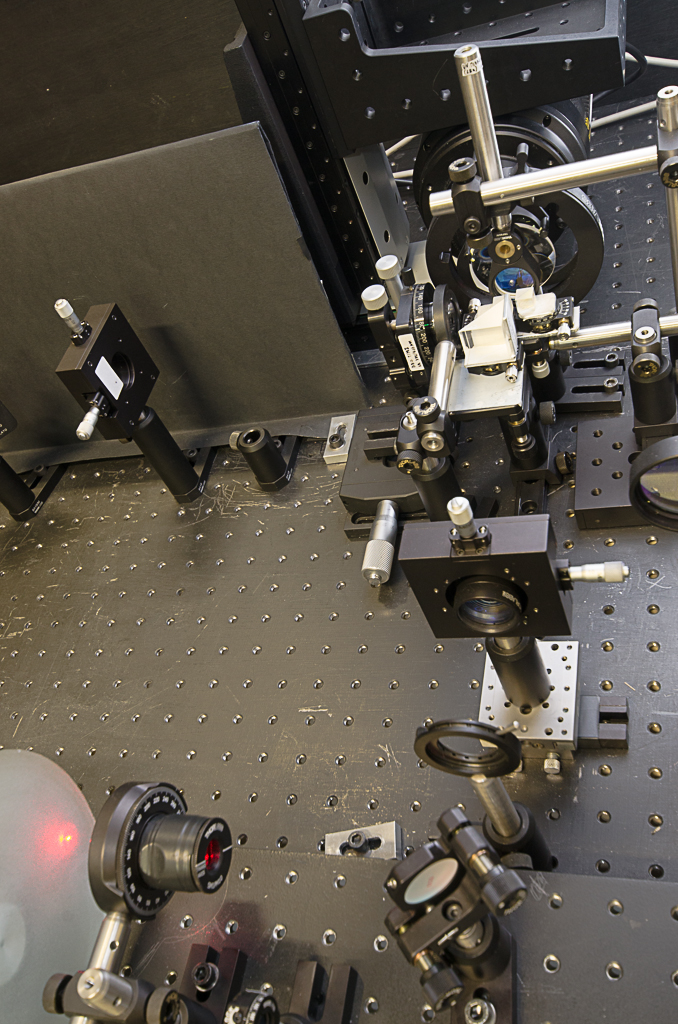}
\caption{Experimental setup. a) Quantum illumination:  b) Classical
  illumination  c) Detected TWB,
  in the presence of the object, without thermal bath. The region of
  interest is selected by an interference filter centered around the
  degeneracy wavelength (710~nm) and bandwidth of 10~nm. After
  selection the filter is removed.  d) Detected field for split
  thermal beams in the presence of the object, without thermal
  bath. e) A typical frame used for the measurement, where the
  interference filter has been removed and a strong thermal bath has
  been added on the object branch. The color scales on the right
  correspond to the number of photons per pixel. On the right hand there is the photo of the set-up.
} \label{setup} \end{center} \end{figure}

In our setup (Figure~\ref{setup}) correlated photon pairs in orthogonal polarisations are generated in Parametric Down Conversion (PDC) process by pumping a BBO (Beta-Barium-Borate) non-linear crystal with the third harmonic (355~nm) of a Q-switched Nd-Yag laser (repetition rate of 10~Hz, 5~ns pulse width) after spatial
filtering. The correlated emissions are then addressed to a high
quantum efficiency (about 80~\% at 710~nm) CCD camera ($N_{\rm pix}=80$ pixels of size
$A_{\rm pix}=(480~\mu\hbox{m})^{2}$). The exposure
time of the camera is set to collect in a single image the emission
generated by a single laser shot. For QI protocol (Figure~\ref{setup}a)
after the BBO crystal, where TWB are generated, one of the beam (the
  ``ancilla'') is reflected towards the detection system. The
  correlated beam is partially detected, together with the thermal
  field from the Arecchi's disk, when the object (actually a beam
  splitter) is present, otherwise it is lost (not showed). Low-pass
  filter (95~\% of transmission at 710nm) and UV-reflecting mirror are used to minimize the background noise while
  maintaining low losses. A lens, placed at the focal length from
  the crystal and the CCD camera, realizes the Fourier transform of
  the field at the output face of the crystal. The PDC light is then combined
at the CCD with a thermal background produced by scattering a laser
beam on an Arecchi's rotating ground glass. When the object is
removed, only the thermal bath reaches the detector. In order to implement CI
protocol (Figure~\ref{setup}b), the TWB are substituted with classical
correlated beams. These are obtained by splitting a multi-thermal beam
(single arm of PDC) and by setting the pump
intensity to ensure equivalent intensity,
time and spatial coherence properties  for the
quantum and the classical sources.

We note that traveling wave PDC generates a spatially
multimode emission in the far field, where each mode corresponds to the transverse component of a specific wavevector. Each pair of correlated modes, corresponding to opposite transverse component of the wavevector with respect to the pump direction, are found in symmetric positions  \cite{Brambilla-PRA-2004}.  Thus, we choose two correlated regions of interests (ROIs) on the CCD array (showed in Figure~\ref{setup} c-d-e). The proper sizing of the pixels and the centering of the 2-dimensional array with sub-mode precision, allows to maximize the collection of the correlated photons for
each pair of pixels and at
the same time to minimize the possible presence of uncorrelated ones  \cite{Pinel:12,pra}.
In our experiment, the correlation in the photon number, even at the quantum level for QI, is
realized independently for each pair of symmetrical (translated)
pixels that belong to the ROIs of the TWB (THB). Therefore, a single image is enough
to evaluate correlation parameters, like covariance, averaging
over the $N_{\rm pix}$ pairs. Albeit not strictly necessary, this is
practically effective because reduces the measurement time (less
images are needed) and avoids to deal with the power instability of the
pump laser from pulse to pulse, which is very destructive in this kind
of application \cite{Bri-OptExp:10}. The number of spatio-temporal modes collected is estimated to be
$M=9\cdot 10^{4}$ by fitting a multithermal statistics. The average number of
PDC photon per mode is $\mu=0.075$. We measured separately the size of
the spatial mode, as the FWHM of the correlation function between the
two beams, $A_{\rm corr}=(120\pm 20\mu\hbox{m})^{2}$. Thus, the number of
spatial modes is about $M_{\rm sp}=A_{\rm pix}/A_{\rm corr}=16\pm 5$ and the number of
temporal modes $M_{\rm t}=M/M_{\rm sp}=(6\pm2)\cdot10^{3}$, the last one being
consistent with the ratio between the pump pulse duration and the
expected PDC coherence time, i.e. 1~ps.

\section{The model of the measurement} \label{The model of the measurement}
In our approach, the ability to distinguish the presence/absence of
the object depends on the possibility of distinguishing between the two
corresponding values of covariance $\Delta_{1,2}$, evaluated experimentally as
\begin{equation} \label{Cov}
\Delta_{1,2}=E[N_{1}N_{2}]-E[N_{1}]E[N_{2}],
\end{equation}
where the quantity $E[X]=
\frac{1}{\mathcal{K}}\sum^{\mathcal{K}}_{k=1} X^{(k)}$ represents the
average over the set of $\mathcal{K}$ realizations corresponding  in our
experiment to the pixels of the ROI, i.e.
$\mathcal{K}=N_{\rm pix}$. Therefore, each image provides a determination of the covariance.
Then we define the signal to noise ratio (SNR) of counting base QI protocol  as the ratio of the mean ``contrast'' to its standard
deviation (mean fluctuation):
\begin{equation} \label{SNRdef}
  f_{\rm SNR}\equiv
\frac{\left|\left\langle \Delta_{1,2}^{\rm (in)}-\Delta_{1,2}^{\rm (out)}\right\rangle\right|}
  {\sqrt{\left\langle\delta^{2}\left(\Delta_{1,2}^{\rm (in)}\right)\right\rangle+\left\langle\delta^{2}\left(\Delta_{1,2}^{\rm (out)}\right)\right\rangle}},
\end{equation}
where ``in'' and ``out'' refer to the presence and absence of the
object, respectively and $\langle \cdots \rangle$ is the quantum
expectation value.  From Eq.~(\ref{Cov}) follows that $\langle
\Delta_{1,2}\rangle = (1-\mathcal{K}^{-1})\langle \delta N_{1}\delta
N_{2}\rangle$ and, for $\mathcal{K}>>1$, $ \langle \delta^{2}
\Delta_{1,2}\rangle\simeq \langle\delta^{2}[\delta N_{1} \delta
N_{2}]\rangle/\mathcal{K}$. These expressions allow calculating  $f_{\rm SNR}$ theoretically.
In particular the denominator can be calculated as

\begin{equation} \label{noise}
\mathcal{K}\left\langle \delta^{2}\Delta_{1,2}\right\rangle\simeq\left\langle \delta^{2}(\delta N_{1}\delta N_{2})\right\rangle\equiv \left\langle \left(\delta N_{1}\delta N_{2}\right)^{2}\right\rangle-\left\langle\delta N_{1}\delta N_{2}\right\rangle^{2}.
\end{equation}

By replacing $\delta N_{2}\mapsto \delta N^{(in)}_{2}+\delta N_{b}$ where  $N^{(in)}_{2}$ is the number
of detected photons that has been reflected by the target, and $N_{b}$ is the uncorrelated background, the
right hand side of (\ref{noise}) can be rewritten as

\begin{eqnarray} \label{noise-approx}
\mathcal{K}\langle \delta^{2}\Delta_{1,2}\rangle&\simeq &  \left\langle \left(\delta N_{1}\delta N^{(in)}_{2}+\delta N_{1} \delta N_{b}\right)^{2}\right\rangle-\left\langle\delta N_{1}\delta N^{(in)}_{2}+\delta N_{1}\delta N_{b}\right\rangle^{2}\\\nonumber
&=&\left\langle \left(\delta N_{1}\delta N^{(in)}_{2}\right)^{2}\right\rangle-\left\langle\delta N_{1}\delta N^{(in)}_{2}\right\rangle^{2}+\left\langle\delta^{2} N_{1}\right\rangle\left\langle\delta^{2} N_{b}\right\rangle\\\nonumber
&=&\left\langle \delta^{2}(\delta N_{1}\delta N^{(in)}_{2})\right\rangle+\left\langle\delta^{2} N_{1}\right\rangle\left\langle\delta^{2} N_{b}\right\rangle,
\end{eqnarray}
where we used the statistical independence of $N_{b}$ and the fact that $\langle\delta N_{b}\rangle=0$.
It is clear that in the absence of the target (situation labeled with the superscript "out"), $N^{(in)}_{2}=0$, thus  $\langle
\delta^{2}\Delta^{(out)}_{1,2}\rangle=\left\langle\delta^{2}
N_{1}\right\rangle\left\langle\delta^{2} N_{b}\right\rangle$, since nothing is reflected to the
detector. However, if the the background fluctuations $\left\langle\delta^{2} N_{b}\right\rangle$
is the largest contribution to the noise, also when the target is present (indicated with superscript "in") we can write $\langle
\delta^{2}\Delta^{(in)}_{1,2}\rangle\simeq\left\langle\delta^{2}
N_{1}\right\rangle\left\langle\delta^{2} N_{b}\right\rangle$. Under this assumption representing a
realistic situation of a very noisy environment, the SNR becomes
\begin{equation} \label{SNR2}
  f_{\rm SNR}\simeq
\frac{\langle \delta N_{1}\delta N_{2}\rangle}
  {\sqrt{2\left\langle\delta^{2} N_{1}\right\rangle\left\langle\delta^{2} N_{b}\right\rangle}}.
\end{equation}
We underline that (\ref{SNR2}) holds for a dominant background, irrespective of its statistics
(e.g. multi-thermal or Poissonian).

In our experiment we consider background with multi-thermal statistics. For a generic multi-thermal
statistics with number of spatiotemporal modes $M$, mean photon number number per mode $\mu$, the
total number of detected photons is $\left\langle N\right\rangle=M \eta \mu$ and the mean squared
fluctuation is $\left\langle\delta^{2} N\right\rangle= M \eta \mu(1+\eta \mu)=\left\langle N
\right\rangle\left(1+\left\langle N\right\rangle/M\right)$ [see for example \cite{MW}, where $\eta$ is the detection
efficiency.

Thus, the amount of noise introduced by the background can be increased by boosting  its total
number of photons $\left\langle N_{b}\right\rangle$ or by varying the number of modes $M_{b}$. 

Moreover, both TWB and correlated THB present locally the same multi-thermal statistics, but with a
number of spatiotemporal modes $M=9\cdot 10^{4}$ much larger than the one used for the background
beam ($M_{b}=57$ in one case and, $M_{b}=1.3 \cdot10^{3}$ in the other). This contributes to make
the condition of preponderant background effective in our realization, even for a relatively small
value of $N_b$.

However, we point out that all the theoretical curves reported in all the Figures are evaluated by
the exact analytical calculation of the four order (in the number of photons) quantum expectation
values appearing on the right hand side of (\ref{noise}), even if the whole expressions are far
more complex than the ones obtained with the assumption of preponderant background.

Starting from (\ref{SNR2}) and considering the same local resources for classical and quantum
illumination beams (in particular the same local variance $\left\langle\delta^{2}
N_{i}\right\rangle_{CI}=\left\langle\delta^{2}  N_{i}\right\rangle_{QI}$ ($i=1,2$)) the enhancement
of the quantum protocol can be easily obtained as

\begin{equation}\label{R}
R=\frac{f^{(QI)}_{\rm SNR}}{f^{(CI)}_{\rm SNR}}\approx\frac{\langle\delta N_{1}\delta N_{2}\rangle_{QI} }{\langle \delta N_{1}\delta N_{2}\rangle_{CI}}= \frac{\varepsilon^{(QI)}}{\varepsilon^{(CI)}}.
\end{equation}
with $\varepsilon = \langle : \delta N_{1} \delta N_{2} : \rangle ~ / ~ \sqrt{\langle :\delta^{2}
N_{1}:\rangle\langle :\delta^{2} N_{2}:\rangle}$ being the generalized Cauchy-Schwarz parameter, where $\langle:~:\rangle$ is the normally ordered quantum expectation value.  This parameter is interesting since it does not depends on the losses and it quantifies non-classicality being $\varepsilon \leq 1$ for classical state of light (with positive $P$-function). The covariance of two correlated beams obtained by splitting a single
thermal beam is $\langle \delta N_{1}\delta N_{2}\rangle_{TH}= M \eta_{1}\eta_{2} \mu_{TH}^{2}$,
while the one of TWB is $\langle \delta N_{1}\delta N_{2}\rangle_{TW}= M \eta_{1}\eta_{2} \mu_{TW}
(1+\mu_{TW})$ (see for example \cite{pra2011GI}). By using this relation with the assumption of the same
local resources, $\mu_{TH}=\mu_{TW}=\mu$ we can derive explicitly $R\approx(1+\mu)/\mu$, which is
insensitive to the amount of noise and loss. On the other side the generalized Cauchy-Schwarz
parameter for a split thermal beam is $\varepsilon_{0}^{(CI)}=1$, where the subscript "$0$" stands
for "in absence of background", as it can be easily derived from the equations of covariance and
single beam fluctuations used previously. Therefore the comparison with split thermal beams represents the comparison with the
"best" classical case.

\section{The results}

\begin{figure}[tbp]
\begin{center}
\includegraphics[width=0.5\textwidth]{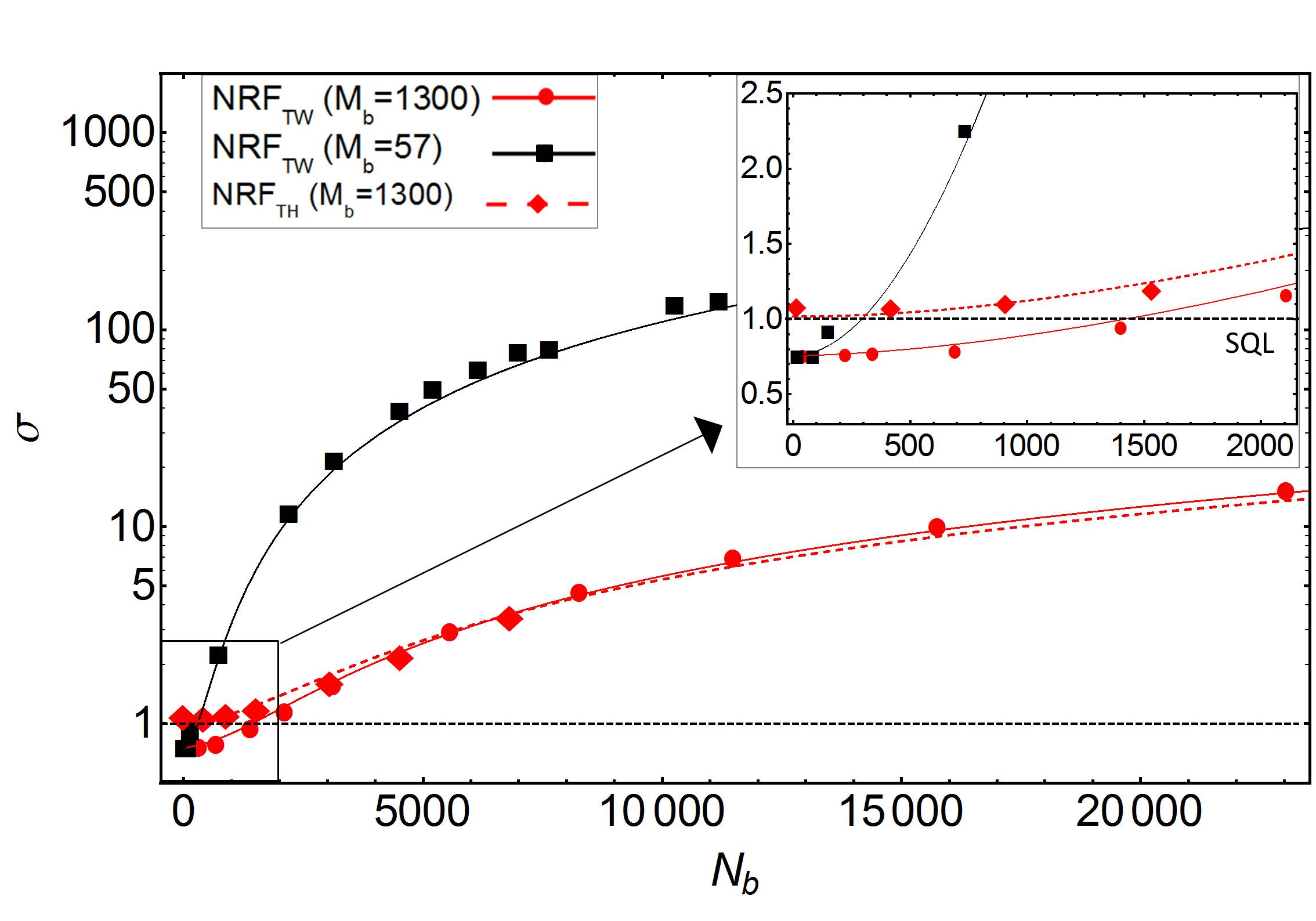}
 \includegraphics[width=0.5\textwidth]{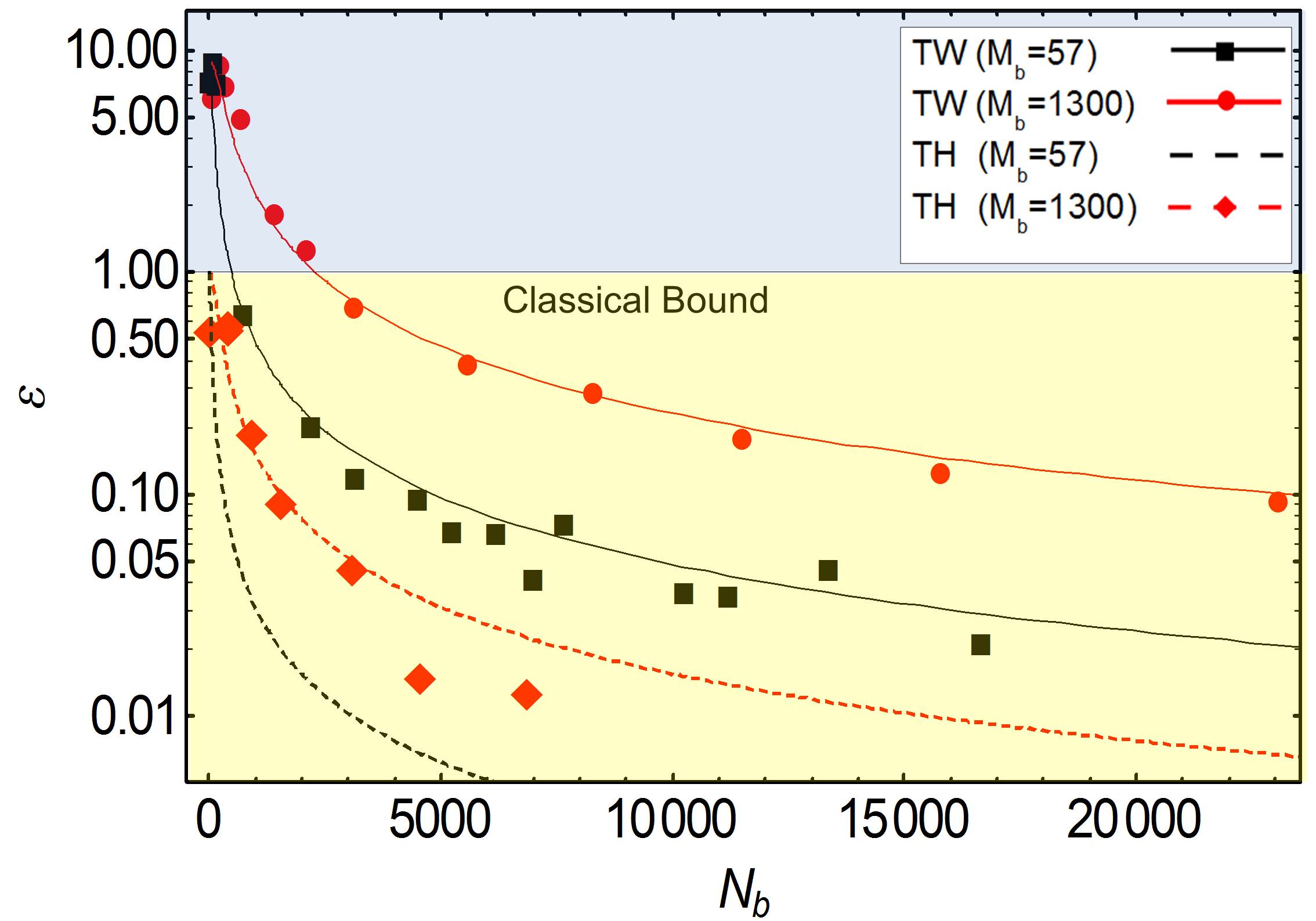}
\caption{Up, NRF in the case of TWB, NRF$_{\rm TW}$, and of the correlated thermal beams, NRF$_{\rm TH}$, as a function of the average number  of background photons $N_{b}$ for $M_{b}=57$ (black series) and
  $M_{b}=1300$ (red). The lines represent the theoretical prediction
  for $\eta_{1}= 2\eta_{2}=0.4$ and $\mu=0.075$ (the last estimated independently). For $N_{b}=0$,
  NRF$_{\rm TW}$ is $\sigma = 0.761\pm 0.006$. Statistical uncertainty bars are too
  small for being visible. Bottom, generalized Cauchy-Schwarz parameter $\varepsilon$  in the case of twin beams, $\varepsilon^{(\rm TW)}$, and of the correlated thermal beams, $\varepsilon^{(\rm TH)}$, as a function of the average number
  of background photons $N_{b}$ for a number of background modes $M_{b}=57$ (black series) and
  $M_{b}=1300$ (red). The lines represent the theoretical prediction at $\mu=0.075$ (the last estimated independently).
 } \label{NRF} \end{center}
\end{figure}

First of all we evaluate the noise reduction factor (NRF) defined as
\cite{bri:10,pra,mas,m}:
\begin{equation}\label{sigma}
\sigma\equiv\frac{\left\langle\delta ^{2}(N_{1}-N_{2})\right\rangle} {\left\langle N_{1}+N_{2}\right\rangle},
\end{equation}
where $\langle N_i \rangle$ is the mean value, and $\delta^2 N_i = (N_i - \langle N_i
\rangle)^2$ is the fluctuation of the photon number $N_{i}$, $i=1,2$, detected by
correlated pixels. It represents the noise of the photon number difference normalized to the shot noise level (SNL) or standard
quantum limit (SQL) \cite{pra}. For classical states $\sigma \ge 1$, while it is always
smaller than 1 for TWB. In particular, when the thermal bath is off, we have $\sigma_{0}=
1-\overline{\eta}+(\eta_{1}-\eta_{2})^{2}\left(1/2+ \mu
\right)/(2\overline{\eta})$, with $\overline{\eta}=(\eta_{1}+\eta_{2})/2$, and $\eta_{i} $ is the
overall detection efficiency of beam $i=1,2$\cite{Bri-OptExp:10,pra}. It includes all the
transmission-detection losses, thus $\eta_{1}= 2\eta_{2}$ due to the presence of the half reflecting object in the path of the second beam.
In Figure~\ref{NRF} we report the measured NRF and the theoretical
prediction. From the inset one can observe that the NRF is actually in
the quantum regime ($\sigma <1$) for small values of
the thermal bath, and in absence of it we obtain $\sigma_{0}=0.76$ corresponding to $\eta_{1}=0.4$.
While, as soon as the contribution of the bath to the fluctuation of $N_{2}$ becomes dominant, NRF increases quite fast well above the classical threshold. As expected from the multi-thermal character of the bath, the number of
modes $M_{b}$ determines the noise level introduced, and it can be tuned easily according to the spin velocity of the
ground-glass disk and/or the acquisition time. We also note that, for THB, the NRF is always in the classical regime.
\par

As a second figure of merit, more appropriate for quantifying the quantum resources exploited by our QI strategy we consider the generalized Cauchy-Schwarz parameter $\varepsilon$ introduced in Sec. \ref{The model of the measurement}. In Figure~\ref{NRF} we report the measured $\varepsilon$ and the theoretical prediction. One observes that for TWB $\varepsilon^{(QI)}$ is actually in the quantum regime ($\varepsilon^{(QI)} >1$) for small values of the thermal background $\langle N_b \rangle $ ($\varepsilon^{(QI)}_{0}\simeq 10$ when $\langle N_b \rangle =0$).  Also here, $\varepsilon^{(QI)}$ decreases quite fast, well below the classical threshold with the intensity of the background. As expected, for THB $\varepsilon^{(CI)}$ is always in the classical regime(being one for $\langle N_b \rangle =0$).

\begin{figure}[tbp]
\begin{center}
 \includegraphics[width=0.5\textwidth]{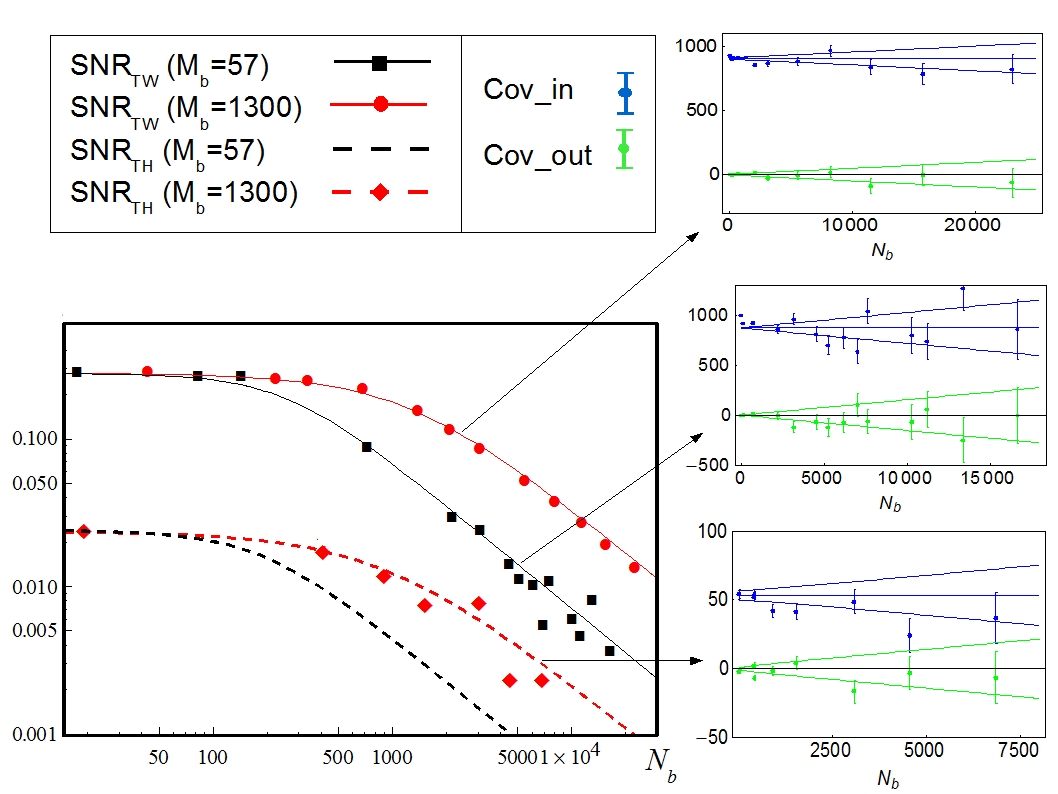}
\caption{SNR versus $N_{b}$ normalized by the square root of number of
  realization. The red (black) markers refer to
  $M_{b}=1300$ ($M_{b}=57$) and the solid (dashed) theoretical curve
  corresponds to quantum (classical) illuminating beams. The lowest
  curve of the classical protocol has not been compared with the
  experimental data because the SNR is so low that a very large number
  of images (out of the possibility of the actual setup) is required
  to have reliable points. The insets on the left present the covariance in the presence, $\Delta_{1,2}^{\rm (in)}$ (blue),
  or absence, $\Delta_{1,2}^{\rm (out)}$ (green), of the target. a)
  and b) refer to QI and CI, respectively, for the same number of bath
  modes $M_{b}=1300$; c) refers to QI with a a lower number of modes,
  $M_{b}=57$. Uncertainty bars represent the uncertainty on the mean values of
  the covariance obtained averaging over the $N_{\rm img}$ images
  (from the top to bottom: $N_{\rm img}=2000,6000$ and
  $4000$). Horizontal lines are the theoretical values
  $\langle\Delta_{1,2}^{\rm (in/out)}\rangle$, while the uncertainty bars
  should be compared with the gap between the dashed lines,
  corresponding to the theoretically evaluated $ \langle \delta^{2}
  \Delta_{1,2}^{\rm (in/out)}\rangle/\sqrt{N_{\rm img}}$.  }
\label{SNR} \end{center} \end{figure}

In Figure~\ref{SNR}, the $f_{\rm SNR}/\sqrt{\mathcal{K}}$ is compared
with the experimental data, where the estimation of quantum mean values of
(\ref{SNRdef}) are obtained by performing averages of $\Delta_{1,2}^{\rm (in/out)}$ over a set of
$N_{\rm img}$ acquired images. While the SNR unavoidably decreases
with the added noise for both QI and CI, the ratio between them is
almost constant ($R\gtrsim 10$) regardless the value of $N_{b}$, in agreement with the results of Sec. \ref{The model of the measurement}.  In
turn, the measurement time, i.e., the number of repetitions $N_{\rm
  img}$ needed for discriminating the presence/absence of the target,
is dramatically reduced (for instance, to achieve $f_{\rm SNR}=1$,
$N_{\rm img}$ is 100 times smaller when quantum correlations are
exploited). Furthermore, Figure~\ref{SNR} shows that the mean value
of the covariance does not depend on the quantity of environmental
noise, because, as expected, only the correlated components survives
to this operation. However, the added noise influences drastically the
uncertainty on the measurement for a certain fixed number of images
$N_{\rm img}$ and thus the ability to assert the presence of the
object.
\par

\begin{figure}[tbp] \begin{center}
 \includegraphics[width=0.5\textwidth]{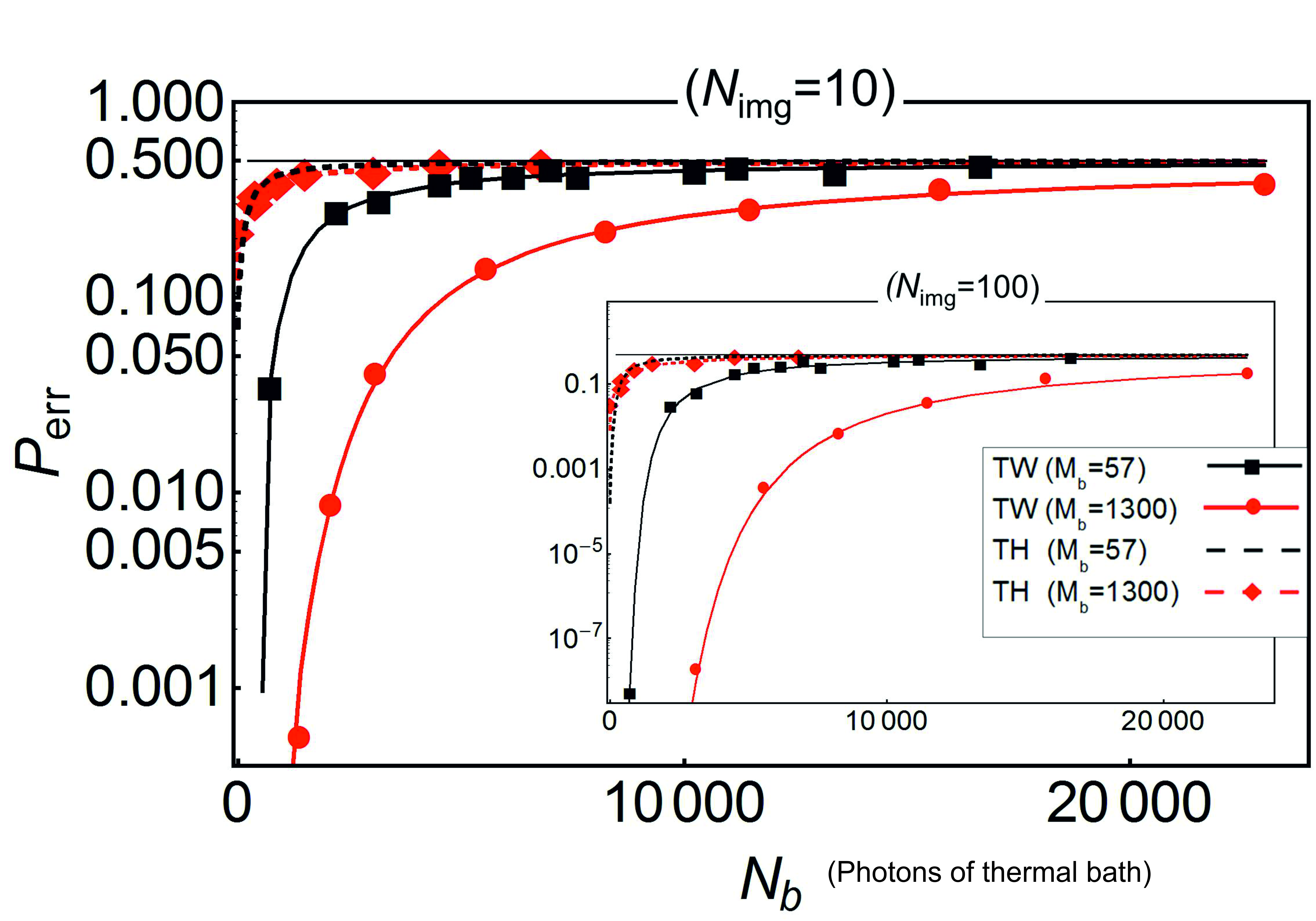}
\caption{Error probability $P_{\rm err}$ of the target detection
  versus the total number of photons of the thermal bath $N_{b}$
  evaluated with $N_{\rm img}=10$ ($N_{\rm img}=100$ in the
  inset). The black squares and red circles are the data for QI with
  $M_{b}=57$ and $M_{b}=1300$, respectively, while red diamonds
  referes to the data for the CI with $M_{b}=1300$. The curves are the
  corresponding theoretical predictions.} \label{Perr} \end{center}
\end{figure}

In order to show that the quantum strategy outperforms the classical
one, in Figure~\ref{Perr} we report the error probability in the
discrimination, $P_{\mathrm{err}}$, versus the number of photons of
the thermal bath $N_{b}$. The statement on the presence/absence of the
object is performed on the basis of the covariance value obtained for
a fixed number of images $N_{\rm img}=10$. Thus, $P_{\mathrm{err}}$ is
estimated fixing the threshold value of the covariance that minimizes
the error probability itself. Figure~\ref{Perr} shows a remarkable
agreement between the theoretical predictions (lines) and the
experimental data (symbols), both for QI and CI
strategy. Furthermore, the $P_{\mathrm{err}}$ in the case of QI is
several orders of magnitude below the CI one and, in terms of
background photons, the same value of the error probability is reached
for a value of $N_{b}$ at least 10 times larger than in the QI case.

\section{Conclusions}

We have described in detail the model and the experiment addressed to quantum enhancement in
detecting a target in a thermal radiation background in a relevant and
realistic measurement scenario. Our system shows quantum correlation
with no external noise ($\sigma=0.76$) even in the presence of the
losses introduced by the only partially reflective target. Remarkably,
even after the transition to the classical regime ($\sigma \gg 1$),
the scheme preserves the same strong advantage with respect its
natural classical counterpart based on classically correlated beams,
as also suggested in \cite{tan:08}. This apparent contradiction is
explained by considering that quantum correlations actually survive
unchanged up to the detector, where they are simply added to a
independent noisy background. \textbf{Moreover the quantum resources and the quantum enhancement achieved by the protocol can be precisely quantified by the generalized Cauchy-Schwarz parameter that is only related to the source properties $\varepsilon$.}
\par
Unlike other quantum enhanced measurement protocols, based on the
experimental estimation of the first-moments of the photon number
distribution, our scheme, which is based on the measurement of the
second order momenta, is impressively robust against losses. This derives from the fact that it
does not require high level of two-mode squeezing ($\sigma_{0}=0.76$ in our experiment). For
instance the quantum imaging protocol \cite{bri:10}, where the signal
is given by $\langle N_{1}-N_{2}\rangle$, provides a maximum improving
factor of $1/\sqrt{\sigma_{0}}$ over classical techniques, that would
correspond to 1.14 in our working condition in the absence of thermal
\textbf{background}. Also in exemplar quantum enhanced schemes, such as detection of
small beam displacement \cite{Treps:03} and phase estimation
by interferometry \cite{gio:11}, it is well known that losses and
noise can rapidly decrease the advantage of using quantum light \cite{Walmsley-PRL-2011}, and
typically high level of squeezing is necessary. This
enforced inside the generic scientific community the common belief that
the advantages of entangled and quantum state are hardly applicable in a
real context, and they will remain limited to proofs of principle
experiments in highly controlled laboratories, and/or to mere academic
discussions. Our work challanges this belief by demonstrating an advantage of orders of magnitude respect to CI protocol,
independently on the amount of thermal noise and using devices available nowadays.
In summary, we believe that
the photon counting based QI protocol has a huge potentiality to
foster the exploitation of quantum light based technologies in real lossy and noisy
environment.

The research leading to these results has received funding from the EU FP7 under grant agreement n.
308803 (BRISQ2), Fondazione SanPaolo and MIUR (FIRB ``LiCHIS'' - RBFR10YQ3H, Progetto Premiale
``Oltre i limiti classici di misura'').

\end{document}